\documentclass[aps,pra,twocolumn,showpacs,floatfix]{revtex4}
\usepackage{graphicx}
\usepackage{graphics}
\begin{document}

\title{Pulse retrieval and soliton formation in a non-standard scheme for dynamic electromagnetically induced transparency}
\author{Amy Peng, Mattias Johnsson, and Joseph J. Hope}
\affiliation{Centre for Quantum Atom Optics, Department of
Physics, The Australian National University, Canberra, ACT 0200,
Australia}

\date{\today}

\begin{abstract}
We examine in detail an alternative method of retrieving the
information written into an atomic ensemble of three-level atoms
using electromagnetically induced transparency. We find that the
behavior of the retrieved pulse is strongly influenced by the
relative collective atom-light coupling strengths of the two
relevant transitions. When the collective atom-light coupling
strength for the retrieval beam is the stronger of the two
transitions, regeneration of the stored pulse is possible.
Otherwise, we show the retrieval process can lead to creation of
soliton-like pulses.
\end{abstract}

\pacs{42.50.Gy, 03.67.-a, 42.65.Tg}

\maketitle

\section{Introduction}
Recent progress in the coherent control of light-matter
interactions has led to many interesting possibilities and
practical applications. Amongst them is the concept of
electromagnetically induced transparency (EIT), first proposed by
Harris \cite{Harris1997}, in which a strong coherent field
(``control") is used to make an otherwise opaque medium
transparent near atomic resonance for a second weak (``probe")
field.  This EIT scheme can be used to store and retrieve the full
quantum information in a weak probe field by changing the strength
of the control field while the pulse is inside the atomic sample
\cite{Fleischhauer2000}. In this paper we examine in detail an
alternative method of retrieving the stored information that was
first proposed by Matsko {\it et al.}~\cite{Matsko2001}, and
investigate the parameter regime under which this process is
feasible.

The usual way of retrieving the stored information consists of a
time-reversed version of the writing process \cite{Liu2001,
Phillips2001}.  The elegant physics behind this scheme was
described by Fleischhauer and Lukin, who noted that the combined
atomic and optical state adiabatically follows a dark state
polariton \cite{Fleischhauer2002}.  The ``writing" process
involves turning the control field to zero, storing the quantum
information of the light beam as the purely atomic form of the
polariton.  When the control field is returned back to its
original value, the polariton switches back to photonic form,
identical to original optical pulse. In work describing some of
the detailed behavior of that process, Matsko {\it et al.} noted
an alternate scheme that may also store and reproduce a copy of a
weak probe pulse \cite{Matsko2001}. In this scheme, the writing
process remains the same, but retrieval of the pulse involves
applying a coherent control field to the transition originally
coupled by the probe field. This causes the probe pulse to be
regenerated on the transition originally coupled by the control
field.  This alternative scheme cannot be explained in terms of
dark state polaritons, and behaves quite differently for different
parameters of the system.  We analyze this scheme in detail in
this paper.

As it is important to distinguish between the different control fields
applied at different times, we will refer to the control field as
the ``writing" field during the first part of the process (storage)
and as the ``retrieval" field during the second part (the
regeneration of the probe pulse).

We will show that when the collective atom-light coupling strength
for the retrieval beam is the largest of the two transitions,
retrieval is possible and the retrieved pulse is amplified, time
reversed, stretched or compressed in time and phase conjugated
compared to the input pulse. Conversely, if the collective
coupling strength of the retrieval beam is not the largest of the
two transitions, we find that the retrieved pulse differs
substantially from the input pulse and the retrieval process can
lead to the creation of a soliton-like combination of
electromagnetic fields and atomic coherences that propagates
without change in shape.

\section{Model}
\begin{figure}
\begin{center}
\includegraphics[width=6cm,height=2.8cm]{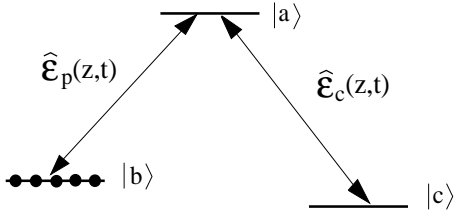}
\caption{Level structure of the atoms} \label{threelevel}
\end{center}
\end{figure}

To analyze the system we use a quasi one-dimensional model,
consisting of two copropagating pulses passing through an
optically thick medium of length $l$ consisting of three-level
atoms. The atoms have two metastable ground states $|b \rangle$
and $|c \rangle$ which interact with two fields
$\hat{\mathcal{E}}_p(z,t)$ and $\hat{\mathcal{E}}_c(z,t)$ as shown
in Figure \ref{threelevel}. $\hat{\mathcal{E}}_i$ are the slowly
varying amplitude related to the positive frequency part of the
electric field given by
\begin{eqnarray}
\hat{E}_p^{+} & = & \sqrt{\frac{\hbar \omega_{ab}}{2 \epsilon_0 V}} \hat{\mathcal{E}}_p (z,t) e^{\frac{i \omega_{ab}}{c}(z-ct)} \nonumber \\
\hat{E}_c^{+} & = & \sqrt{\frac{\hbar \omega_{ac}}{2 \epsilon_0
V}} \hat{\mathcal{E}}_c (z,t) e^{\frac{i \omega_{ac}}{c}(z-ct)}.
\nonumber
\end{eqnarray}
Here $\omega_{\mu \nu} = (E_\mu - E_\nu)/\hbar$ is the resonant
frequency of the $|\mu \rangle \leftrightarrow | \nu \rangle$
transition and $V$ the quantization volume, here taken as the
interaction volume. As the pulses are
co-propagating we are able to neglect Doppler effects.

To perform a quantum analysis of the light-matter interaction it
is useful to use locally-averaged atomic operators. We take a
length interval $\delta z$ over which the slowly-varying field
amplitudes do not change much, containing $n \mathcal{A} \delta z
\gg 1$ atoms, where $n$ is the atomic density and $\mathcal{A}$ is
the cross sectional area of the pulses, and introduce the
continuous atomic operators
\begin{equation}
\hat{\sigma}_{\mu \nu}(z,t) = \frac{1}{n \mathcal{A} \delta z}
\sum_{i, z \leq z_i < z + \delta z} | \mu^i(t) \rangle \langle
\nu^i(t) | e^{ \frac{i \omega_{\mu \nu}}{c}(z - ct)}.
\label{contatom}
\end{equation}
where $z_i$ is the position of the $i$th atom and $| \mu^i(t)
\rangle$ is the $| \mu \rangle$ state wavefunction for the $i$th
atom..

Using these continuous atomic operators, the interaction
Hamiltonian under the rotating wave approximation is given by
\begin{equation}
\hat{\mathcal{H}} = - \int_0^{l} \frac{N \hbar}{l} [ g_p
\hat{\mathcal{E}}_p(z,t) \hat{\sigma}_{ab}(z,t) + g_c
\hat{\mathcal{E}}_c(z,t) \hat{\sigma}_{ac}(z,t) + H.c.] dz
\label{hamiltonian}
\end{equation}
where $l$ is the length of the cell, $N$ is the number of atoms in
the interaction region and the coupling constants are $g_p =
d_{ab} \sqrt{\omega_{ab}/2 \epsilon_0 V \hbar}$, $g_c = d_{ac}
\sqrt{\omega_{ac}/2 \epsilon_0 V \hbar}$ where $d_{ab}$ and
$d_{ac}$ are the dipole moments of the $|a \rangle \leftrightarrow
| b \rangle$ and $| a \rangle \leftrightarrow |c \rangle$
transitions respectively. This Hamiltonian leads to the following
equations of motion for the density matrix elements and fields
\begin{eqnarray}
\dot{\rho}_{bb} & = & \gamma_b \rho_{aa} - i \Omega_p \rho_{ba} + i\Omega_p^{\ast} \rho_{ab} \label{rhobb} \\
\dot{\rho}_{cc} & = & \gamma_c \rho_{aa} - i \Omega_c \rho_{ca} +
i \Omega_c^{\ast} \rho_{ac} \nonumber \\
\dot{\rho}_{ab} & = & - \gamma_{ab} \rho_{ab} + i \Omega_p
(\rho_{bb} - \rho_{aa}) + i \Omega_c \rho_{ab} \nonumber \\
\dot{\rho}_{cb} & = & - i \Omega_p \rho_{ca} + i \Omega_c^{\ast}
\rho_{ab} \nonumber \\
\dot{\rho}_{ca} & = & - \gamma_{ca} \rho_{ca} - i \Omega_p^{\ast}
\rho_{cb} + i \Omega_c^{\ast}( \rho_{aa} - \rho_{cc} )
\nonumber \\
\left( \frac{\partial}{\partial t} \right. & + & \left. c
\frac{\partial}{\partial z} \right) \Omega_p = i \alpha_p
\rho_{ab} \nonumber \\
\left( \frac{\partial}{\partial t} \right. & + & \left. c
\frac{\partial}{\partial z} \right) \Omega_c = i \alpha_c
\rho_{ac} \label{motion}
\end{eqnarray}
where the $\gamma_\mu$ are phenomenological decay rates and we
have defined the two Rabi frequencies as $\Omega_i = g_i \langle
\hat{\mathcal{E}}_i \rangle$ and the constants \mbox{$\alpha_p = N
|g_p|^2$} and \mbox{$\alpha_c = N |g_c |^2$} are the collective
atom-light coupling constants for the transitions $| b \rangle
\leftrightarrow | a \rangle$ and $| c \rangle \leftrightarrow | a
\rangle$ respectively.

We now proceed to consider the modified version
\cite{Matsko2001,Zibrov2002} of the storage and retrieval scheme
using dynamic EIT. With all the atoms initially optically pumped
into the ground state $| b \rangle$, we turn on the writing beam
driving the $| c \rangle \leftrightarrow | a \rangle$ transition
to its maximum value $\Omega_c^0$, while a small input pulse of
maximum amplitude $\Omega_p^0 \ll \Omega_c^0$ is sent into the
medium on the $| b \rangle \leftrightarrow | a \rangle$
transition. EIT effects cause the group velocity of the the input
pulse to be drastically reduced to a new value given by
\cite{Fleischhauer2002}
\begin{equation}
v_g = \frac{c}{ 1 + \alpha_p / |\Omega_c^0|^2} \label{vg}
\end{equation}
which can be much less than the speed of light. This leads to
significant spatial compression of the input pulse as it enters
the medium, allowing the entire pulse (the typical pulse length
outside the cell is of order of a few kilometers) to be stored
inside a vapor cell (of length a few centimeters). Once the input
pulse is completely inside the medium, we slowly turn off the
writing beam on the $| c \rangle \leftrightarrow | a \rangle$
transition. This writes the information carried by the input pulse
onto a collective atomic coherence for storage
\cite{Fleischhauer2002}.

After a controllable storage time $T_s$, the stored information
can be retrieved by turning on a retrieval beam driving the $| b
\rangle \leftrightarrow | a \rangle$ transition as proposed in
\cite{Matsko2001,Zibrov2002}. Note that this is in contrast to the
usual retrieval scheme for EIT light storage in which the
retrieval beam is on the $| c \rangle \leftrightarrow | a \rangle$
transition and the output pulse is generated on the $| b \rangle
\leftrightarrow | a \rangle$ transition \cite{Fleischhauer2002}.
For normal EIT in the ideal case, the output pulse is identical to
the input pulse. Here, because the time reversed version of the
writing beam is on the $| b \rangle \leftrightarrow |a \rangle$
transition, the output pulse is generated on the $| c \rangle
\leftrightarrow | a \rangle$ transition. As a consequence the
output pulse need not have the same frequency or polarization as
the input pulse.

It is not immediately clear why any fields made by this process would be
correlated with the stored pulse. After all, the original pulse produces a small population in state $|c\rangle$ and leaves most atoms in state $|b\rangle$, so the control beam is mainly interacting with states that have not been affected by the original pulse in any significant way.   Also, as the strong retrieval
beam is initiated on the most populated transition, it will lead
to significant spontaneous emission, which would appear to dominate any kind of coherent process necessary for the
retrieval. However, as will be shown later, our simulation
indicates that under some parameter regimes, it is indeed possible
to recover an output related in amplitude and phase to the input
pulse.

\section{Retrieving the probe pulse}

\begin{figure}
\begin{center}
\includegraphics[width=9cm,height=8cm]{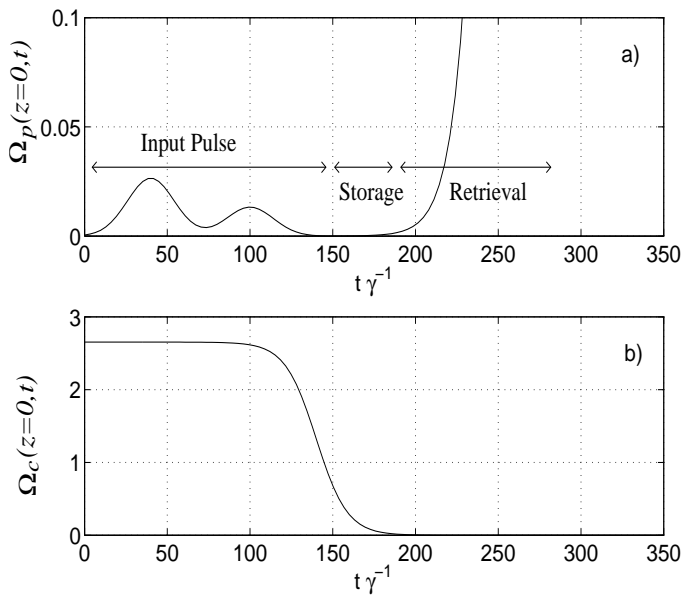}
\caption{Envelope of pulses entering the medium at $z=0$ as a
function of time. a) $\Omega_p$ is the field driving the $| b
\rangle \leftrightarrow | a \rangle$ transition and b) $\Omega_c$
is the field driving the $| c \rangle \leftrightarrow |a \rangle$
transition. The parameters are $T_{\mbox{\it off}} = 140
\gamma^{-1}$, $T_s = 18.85 \gamma^{-1}$, $T_{on} = 259
\gamma^{-1}$, $\Omega_p^0 = 0.0265 \gamma$ and $\Omega_c^0 =
2.6526 \gamma$. Amplitude of pulses are displayed in units of
$\gamma$. The input pulse to be stored is $\Omega_p$ before the
start of storage. The retrieved pulse is $\Omega_c$ after the
storage process.} \label{inputpulses}
\end{center}
\end{figure}

We will start with an example that illustrates a successful pulse
retrieval, and then go on to analyze other possible behaviors of
this system. We assume standard atomic initial conditions for EIT
given by $\rho_{bb}(z,0)=1$; $\rho_{cc}(z,0) = \rho_{ab}(z,0) =
\rho_{cb}(z,0) = \rho_{ca}(z,0) = 0$. In order to better identify
the properties of the new retrieval scheme and to judge the
quality of the retrieval process, we choose an input pulse at
$z=0$ to be the sum of two Gaussians of different height (so the
total input pulse is non-symmetric) with a time-varying phase
factor. The boundary conditions for the fields at $z=0$ is of the
form
\begin{eqnarray}
\Omega_p(0,t) & = & \Omega_p^0 f(t) e^{i f(t)}  +
\frac{\Omega_c^0}{2} \left[ 1 + \tanh \left( \frac{t -
T_{on}}{T_s}
\right) \right]  \label{omegapin} \\
\Omega_c(0,t) & = & \frac{\Omega_c^0}{2} \left[ 1 - \tanh \left(
\frac{t-T_{off}}{T_s} \right) \right] \label{omegacin}
\end{eqnarray}
where $f(t)$ is a unit amplitude envelope function. The first part
of equation (\ref{omegapin}) describes the input (signal) pulse
that we wish to store, whereas the second part of equation
(\ref{omegapin}) describes the turning on of the retrieval beam
(amplitude $\Omega_c^0$) at time $T_{on}$ with a switching time of
approximately $T_s$. $\Omega_c$ describes the turning off of the
writing beam at time $T_{\mbox{\it off}}$. These boundary
conditions for the fields are plotted in Figure \ref{inputpulses}.

With these initial conditions, the equations of motion
(\ref{motion}) can be solved numerically in the moving frame
defined by $\xi = z$, $\tau = t - z/c$ using the method described
in Shore \cite{Shore} implementing the numerical integration with
a fourth order Runge-Kutta method \cite{XMDS}. For convenience, we
choose $\gamma_b = \gamma_{ab} = \gamma$ and $\gamma_c =
\gamma_{ca} = (\alpha_c/\alpha_p) \gamma$.

\begin{figure}
\begin{center}
\includegraphics[width=8cm,height=8cm]{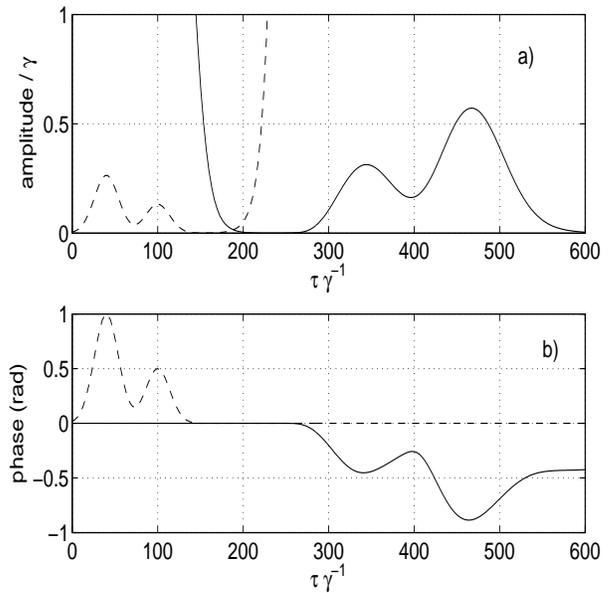}
\caption{Comparison of a) amplitude and b) phase of both the
retrieved and input pulses. The solid line represents the
retrieved pulse and the dashed line the input pulse (magnified by
a factor of ten). The parameters are $\alpha_p = 30177 \gamma$,
$\alpha_c = 29272 c\gamma$, $l = 4$cm. The input pulses are as
shown in Figure \ref{inputpulses}. The large amplitude on the
retrieval pulse transition at earlier times is because the field
on this transition was used as the writing beam initially.
Similarly the large value on the input pulse transition at later
times represents the retrieval beam.} \label{result1}
\end{center}
\end{figure}

Figure \ref{result1} shows a comparison of the phase and amplitude
of the retrieved and input pulses, both propagating in the
positive $z$ direction. In practice, this output pulse of
well-defined shape will be superimposed on top of spontaneously
emitted photons, so for the purposes of observing this output, it
would be better to choose the lowest atomic density allowed for
EIT to work.

\begin{figure}
\begin{center}
\includegraphics[width=8cm,height=8cm]{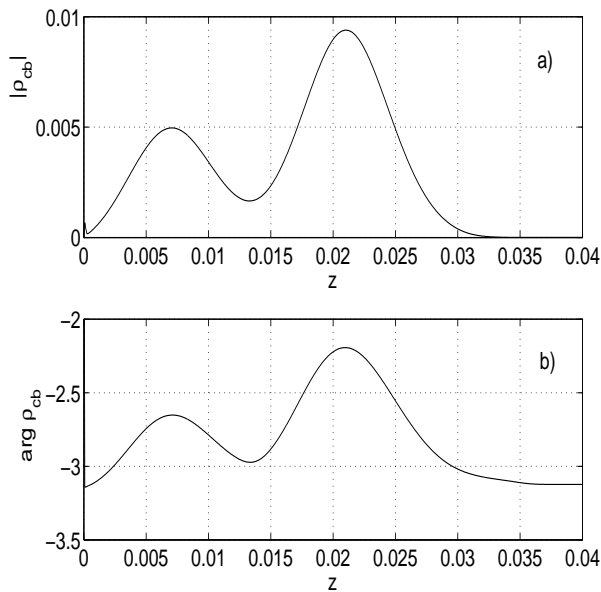}
\caption{Amplitude a) and phase b) of $\rho_{cb}$ during storage.}
\label{stored}
\end{center}
\end{figure}

From the graphs, we observe that the retrieved pulse is time
reversed, amplified, widened in time and is the phase conjugate of
the input pulse. We can understand this behavior by examining the
interaction of the retrieval beam with the stored coherence and
how the retrieved pulse is generated. As the writing part of our
process is identical to the usual dynamic EIT setup
\cite{Fleischhauer2002}, we know that at the end of the storage
process, the only variable of the system aside from $\rho_{bb}$
that is significantly nonzero is $\rho_{cb}$ whose spatial
variation, shown in Figure \ref{stored}, encodes the phase and
amplitude information of the original input pulse.

\begin{figure}
\begin{center}
\includegraphics[width=8cm,height=3.5cm]{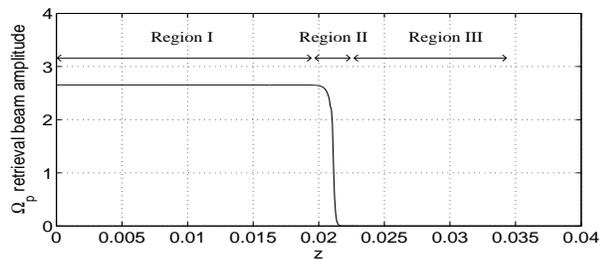}
\caption{Spatial variation of the retrieval beam divides the
medium into three regions as described in the text. At any time,
the retrieved pulse is generated due to the dynamics that occur in
region II where the retrieval beam is in the process of pumping
population out of state $| b \rangle$. In region I, all atoms are
in state $|c \rangle$. The beam has not yet penetrated to region
III.} \label{RetrievalBeam}
\end{center}
\end{figure}

The retrieval beam enters the medium driving the transition $| b
\rangle \leftrightarrow | a \rangle$ while almost all the atoms
are in state $| b \rangle$. As the medium is optically thick, the
retrieval beam is strongly absorbed and its initial wavefront
moves across the medium at a speed much less than the speed of
light. The spatial variation of the retrieval beam at any time
during the retrieval process has the general shape shown in Figure
\ref{RetrievalBeam}. At any time, this shape divides the medium
into three regions with distinct dynamics. In region I, the
retrieval beam $\Omega_p$ has attained its maximum value as it has
optically pumped all atoms into the state $| c \rangle$. In region
II, the wavefront of the retrieval beam is severely attenuated due
to absorption and in region III, not yet reached by the retrieval
beam, all the atoms are still in the polariton state left by the
writing process. This means that most of the atoms are in state $|
b \rangle$ here. We note that at any point in time, the
interesting dynamics related to the retrieval of the stored
information is occurring only in region II. In this region the
retrieval beam $\Omega_p$ is in the process of pumping atoms from
$| b \rangle$ to $| a \rangle$ at a rate of $| \Omega_p |^2 /
\gamma$. Before a significant population has accumulated in $| c
\rangle$, the retrieval beam can coherently scatter off the stored
coherence $\rho_{cb}$, generating a retrieved pulse on the $| a
\rangle$ to $| c \rangle$ transition. These generated photons will
encode some of the properties of the input pulse and propagate out
of the medium at the speed of light through region III, as this
region contains only atoms in $| b \rangle$.

On long time-scales optical pumping alters $\rho_{cb}$ and will
overwrite any coherence that was initially stored there. At this
point, photons emitted when atoms decay from $| a \rangle$ to $| c
\rangle$ will bear no relation to the input pulse. Furthermore,
any field on the $| c \rangle \leftrightarrow | a \rangle$
transition will be strongly damped due to the significant
population in state $| c \rangle$.

\begin{figure}
\begin{center}
\includegraphics[width=8cm,height=7cm]{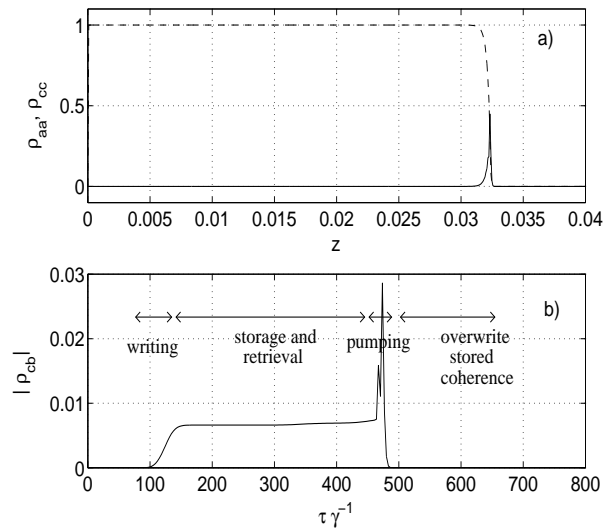}
\caption{a) $\rho_{cc}$ (dashed) and $\rho_{aa}$ (solid) as a
function of $z$ during the retrieval process. A significant
population resides in the excited state during the retrieval
process. b) $|\rho_{cb}(0.024,\tau)|$ as a function of $\tau
\gamma^{-1}$. The spike in the graph shows the effect of optical
pumping by the retrieval beam which overwrites the stored
coherence.} \label{popa}
\end{center}
\end{figure}

Note that in contrast to the adiabatic following in the usual EIT
procedure \cite{Fleischhauer2002}, the retrieval process here is
non-adiabatic and as demonstrated in Figure \ref{popa}(a), the
population in the excited state $| a \rangle$ can be quite
significant during the retrieval process.

Using this description of the dynamics, one can explain many of
the features exhibited by the retrieved pulse. The time reversal
occurs because the wavefront of the retrieval beam moves slowly
from left to right, so the part of the stored coherence that is
closest to the cell entrance sees the retrieval beam first. This
part of the coherence near the medium entrance corresponds to the
tail end of the input pulse (i.e. the part of the input pulse that
entered the cell last), so the tail end of the input pulse will be
retrieved first, resulting in an output that is time reversed.
Figure \ref{result1} also demonstrates that the retrieved pulse
can be amplified, and in this particular example the amplitude is
increased by a factor of about twenty. This occurs because the new
retrieval scheme has a larger reservoir of atoms available for
producing photons. In the normal EIT setup photon number in the
retrieved pulse is limited by the number of atoms in state $| c
\rangle$ during the storage process which is strictly less than
the number of photons in the input pulse. In this scheme the
number of photons generated is limited by the maximum number of
atoms in state $|c \rangle$ that the medium can sustain without
destroying the stored coherence. This is a property of both the
size of the stored coherence and the medium (e.g. optical
density). It is, however, independent of the properties of the
retrieval beam. For an optically dense medium, this is generally
much larger than the number of photons in the input pulse.

\begin{figure}
\begin{center}
\includegraphics[width=8cm, height=7cm]{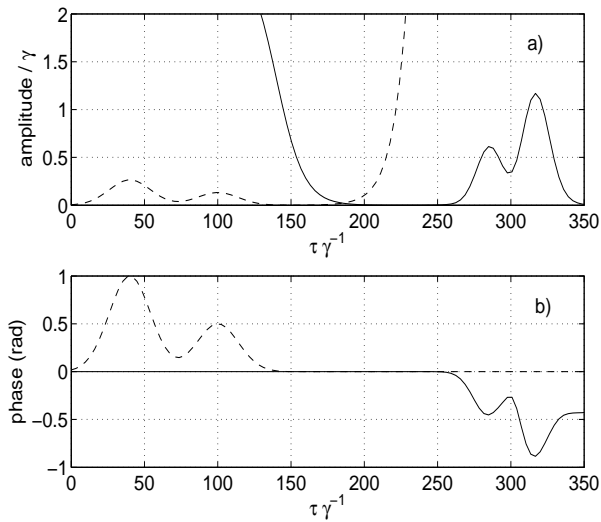}
\caption{Comparison of a) amplitude and b) phase of retrieved
(solid) and input (dashed) pulses. The input pulse has been
magnified by a factor of ten. The amplitude of the retrieval beam
is now $\Omega_p^0 = 5.3052 \gamma$ twice the value in Figure
\ref{result1}. We see that the retrieved pulse is made earlier in
time and narrower in time because of the increased pumping rate by
the retrieval beam. All other parameters are identical to those in
Figure \ref{result1}.} \label{result4}
\end{center}
\end{figure}
It is also clear that the width of the retrieved pulse is
determined by the speed at which the wavefront of the retrieval
beam propagates across the medium and not the initial pulse
length. Figure \ref{result4} shows the retrieved pulse
when the amplitude of the retrieval beam is doubled compared to
Figure \ref{result1} while keeping all other parameters identical.
We see that the output pulse is narrower and its amplitude larger
while the total number of photons (as indicated by the area
underneath the graph of intensities) remains approximately the
same. This is due to the the more intense retrieval beam now being
able to pump atoms out of $| b \rangle$ at a higher
rate, enabling it to move across the medium more quickly and
ensuring a shorter time interval between the retrieval of the
front and back part of the original input pulse. Since the total
number of generated photons is unaffected by the intensity of the
retrieval beam, energy conservation necessitates that the
retrieved pulse has greater amplitude.

The phase conjugation of the output pulse is most easily
understood by examining the equations of motion (\ref{motion}) and
(\ref{rhobb}). On a short time scale, we have $\rho_{bb} \approx 1
\gg \rho_{cc}, \rho_{aa}$, and the following set of equations
describes the generation of the retrieved pulse from the conjugate
of the stored coherence $\rho_{cb}^{\ast}$
\begin{eqnarray}
\left( \frac{\partial}{\partial t} \right. & + & \left. c
\frac{\partial}{\partial z} \right) \Omega_c = i \alpha_c
\rho_{ac} \label{outputeq} \\
\rho_{ac}(z,t) & = & i \int_{T_{on}}^t e^{- \gamma (t-s)}
\Omega_p(z,s) \rho_{cb}^{\ast}(z,s) ds \label{rhoacexp}.
\end{eqnarray}
Since $\rho_{cb}^{\ast}$ stores the conjugate of the input phase,
the retrieved pulse is therefore the phase conjugate of the input
pulse.

\begin{figure}
\begin{center}
\includegraphics[width=8cm,height=7cm]{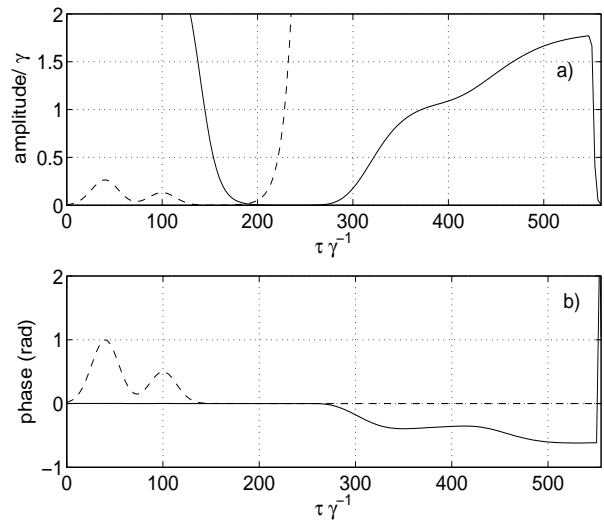}
\caption{Comparison of a) amplitude and b) phase of retrieved
(solid) and input (dashed) pulses. The input pulse
has been magnified by a factor of ten. The abrupt
cut off of the retrieved pulse near $\tau \gamma^{-1} = 600$ is
due to the arrival of the retrieval beam at the exit of the cell
$z=0.04$ which pumps all atoms into state $| c \rangle$ and prevents
any field existing on this transition. The parameters are
$\alpha_p = \alpha_c = 30177 \gamma$, $l =
4$cm. The input pulses are as given in Figure \ref{inputpulses}}
\label{result2}
\end{center}
\end{figure}

\begin{figure}
\begin{center}
\includegraphics[width=8cm,height=7cm]{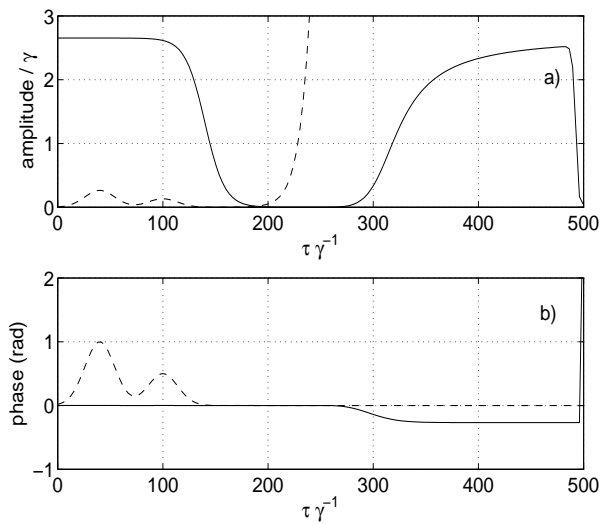}
\caption{Comparison of a) amplitude and b) phase of retrieved
(solid) and input (dashed) pulses. The input pulse has been
magnified by a factor of ten. The parameters are $\alpha_p = 30177
c \gamma$ < $\alpha_c = 31082 c \gamma$, $l = 4$cm. The input
pulses are as given in Figure \ref{inputpulses}.} \label{result3}
\end{center}
\end{figure}

From examining the phase of the generated output field compared to
the input, it is clear that the quality of the retrieval process,
(in terms of extracting a pulse of the same shape as the time
reversed input) is not perfect. For example, from Figure
\ref{result1} (b), we see that the phase of the retrieved pulse
fails to decay to zero after a certain time. This feature is even
more prominent when we set the collective atom-light coupling
constants to be equal $\alpha_p = \alpha_c$, as shown in Figure
\ref{result2}. Again we see that the phase of the retrieved pulse
plateaus, this time near the peak of the second Gaussian.
Increasing $\alpha_c$ further to move into the regime $\alpha_c >
\alpha_p$, we see from Figure \ref{result3} that the quality of
the retrieval process is so bad that the output pulse does not
even display the characteristic double peak of the input pulse.

\begin{figure}
\begin{center}
\includegraphics[width=8cm,height=7cm]{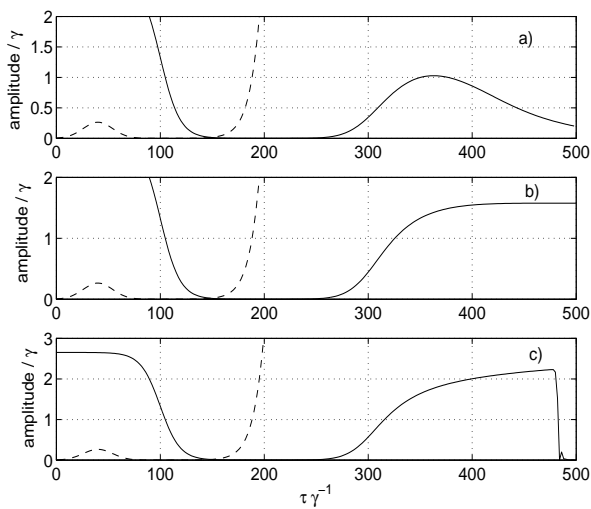}
\caption{Comparison of amplitudes of input pulse (dashed)
magnified by a factor of ten and output pulse (solid) for the
three parameter regimes $\alpha_p>\alpha_c$ (a),
$\alpha_p=\alpha_c$ (b) and $\alpha_p<\alpha_c$.} \label{gaussamp}
\end{center}
\end{figure}

To further investigate the behavior of the retrieved pulse as the
relative strength of the collective coupling constant is varied,
we consider a Gaussian input envelope as this makes clearer the
difference in retrieval quality. Figure \ref{gaussamp} show the
three cases $\alpha_p>\alpha_c$, $\alpha_p = \alpha_c$, and
$\alpha_p< \alpha_c$.
For $\alpha_p > \alpha_c$, the generation of the output pulse
automatically ceases so that the retrieved pulse will always
display the falling end of the Gaussian input. However, as
$\alpha_c$ approaches the value of $\alpha_p$, the pulse
generation takes a longer time to cease giving the assymetric long
tail shape shown in Figure \ref{gaussamp} (a). When $\alpha_p =
\alpha_c$, eventually a constant value of the generated field is
maintained until the retrieval beam reaches the exit of the cell
and pumps all the atoms to $| c \rangle$. Continuing the trend,
Figure \ref{gaussamp} (c) shows that in the regime of $\alpha_p <
\alpha_c$, the generated pulse amplitude grows until the retrieval
beam has optically pumped all the atomic population into $| c
\rangle$. The phase variation of the retrieved pulse confirms the
general trend that quality of the retrieval process decreases as
we move across the three parameter regimes.
\section{Soliton-Like Behavior}
Our numerical results strongly indicate the existence of a soliton
solution at the critical point $\alpha_p = \alpha_c$. At this
point the field $\Omega_c$ generated from the stored coherence is
able to induce a nonzero coherence $\rho_{cb}$ by a co-operative
action with the retrieval beam $\Omega_p$. The induced $\rho_{cb}$
in turn can then be used to generate $\Omega_c$. This steady state
cycling action explains why the system can continue to generate an
output even when the 'dynamic' region (region II in Figure
\ref{RetrievalBeam}) has moved beyond where the input pulse was
originally stored. More specifically, when $\alpha_p > \alpha_c$,
the induced coherence is not large enough to maintain the cycling
action, causing the effect to die out; if $\alpha_p = \alpha_c$,
the effect is self sustaining where the induced $\rho_{cb}$ is
just sufficient to maintain the value of $\Omega_c$ that generated
it; and for $\alpha_p<\alpha_c$, a greater coherence ($\rho_{cb}$)
is induced, leading to an output at the cell exit that is
continually amplified in time until cut off by the arrival of the
retrieval beam. However, we have observed that when the cell
length is increased, the amplitude of the output pulse appears to
tend to a limiting value.

\begin{figure}[t]
\begin{center}
\includegraphics[width=8cm,height=7cm]{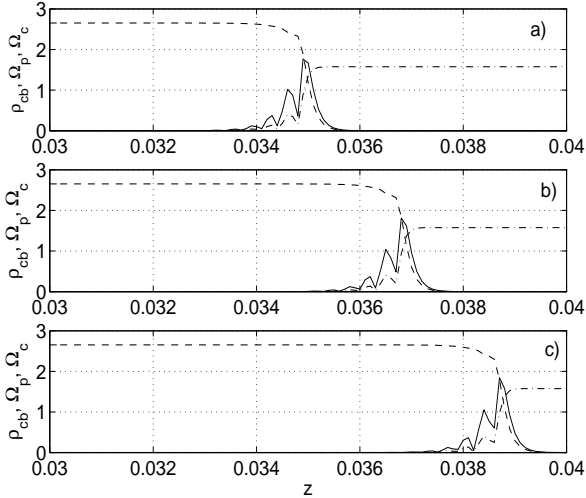}
\caption{$|\Omega_p|$ (dash), $|\Omega_c|$ (dash-dot) and
$|\rho_{cb}| \times 5$ (solid) as a function of $z$ at times $\tau
\gamma^{-1} = 477.6$ (a), $489$ (b), $501.6$ (c) for $\alpha_p =
\alpha_c$ case. The field and atomic coherence propagates together
unchanged across the cell} \label{sameRcb}
\end{center}
\end{figure}

Figure \ref{sameRcb} shows the induced $\rho_{cb}$, $\Omega_p$ and
$\Omega_c$ for $\alpha_p = \alpha_c$ when the initially stored
coherence no longer exists. We note that the shape and size of
$\rho_{cb}$ and the fields propagate unchanged across the cell,
indicating behavior characteristic of solitons. We also found that
the shape and size of the final output is independent of the
values of the coherences originally stored, indicating that the
final field-coherence formation is a characteristic of the system
independent of the storage/retrieval process requiring an initial
nonzero $\rho_{bc}$ coherence only as a seed.

To demonstrate analytically that our results are in fact solitons
we substitute the ansatz $\Omega_i(z,t) = \Omega_i(z-vt)$,
$\rho_{\mu \nu}(z,t) = \rho_{\mu \nu}(z-vt)$ into equations
(\ref{motion}) for the case $\alpha_p = \alpha_c$, where $v$ is
the soliton parameter that designates the constant speed with
which the soliton propagates across the medium. This allows us to
determine the following relationship between the two pulses and
the coherence generated as
\begin{equation}
\rho_{cb}(s) = - \frac{c - v}{\alpha_0 v} \Omega_p(s)
\Omega_c^{\ast}(s) \label{coherencegen}
\end{equation}
where $\alpha_p = \alpha_c = \alpha_0$ is the collective
light-atom coupling constant identical for both transitions and
$s=z-vt$. We also obtain the following relationship between the
limiting values of the two fields
\begin{equation}
\frac{c-v}{\alpha_0 v} (|\Omega_p^{\infty}|^2 +
|\Omega_c^{\infty}|^2) = 2 \label{limitvalues}
\end{equation}
where $ \Omega_p^{\infty} = \lim_{s \rightarrow -\infty}
\Omega_p(s)$ and $ \Omega_c^{\infty} = \lim_{s \rightarrow \infty}
\Omega_c(s)$. When we compare equations (\ref{coherencegen}) and
(\ref{limitvalues}) against the results of our numerics for
various soliton parameters (the soliton parameter can be changed
by varying the amplitude of the retrieval beam), we find good
agreement. After further elimination of all the atomic variables,
we obtain the two soliton equations relating the two fields
\begin{eqnarray}
v \left[ \Omega_p \frac{d^2 \Omega_c^{\ast}}{ds^2}  -
\Omega_c^{\ast} \frac{d^2 \Omega_p}{ds^2} \right]   -  \gamma
\left[\Omega_p \frac{d \Omega_c^{\ast}}{ds} \right. & - & \left.
\Omega_c^{\ast}
\frac{d \Omega_p}{ds} \right] \nonumber \\
 =  - \Omega_p \Omega_c^{\ast} \left( \frac{|\Omega_p^{\infty}|^2}{v}
\right. & - & \left. \frac{\alpha_0}{c-v} \right)
\end{eqnarray}
\begin{eqnarray}
 - \Omega_p \left( v \frac{d^3 \Omega_p}{ds^3}  -   \gamma
\frac{d^2 \Omega_p}{ds^2} \right)  & + &  \left(\frac{d
\Omega}{ds} + \frac{2 \gamma \Omega_p}{v} \right) \left(v
\frac{d^2 \Omega_p}{ds^2} -
\gamma \frac{d \Omega_p}{ds} \right) \nonumber \\
& = & \frac{2 \Omega_p^2}{v} \frac{d}{ds} \left(|\Omega_c|^2 +
|\Omega_p|^2 \right) \nonumber \\
& + &  \frac{\gamma}{v^2} \Omega_p^2(|\Omega_p^{\infty}|^2 -
|\Omega_c|^2 - |\Omega_p|^2 ) \label{soliton2}
\end{eqnarray}
from which the dispersive and nonlinear terms are clearly visible.

We believe the photons generated from the cycling action outlined
above bear little relation to the input pulse and are therefore
not useful as far as quantum information retrieval is concerned.
However the existence of the soliton-like solution and its
sensitivity with respect to the coupling parameters is likely to
lead to other interesting possibilities. A further analysis of the
generation and properties of the solitons is beyond the scope of
this paper, although the possibility that solitons can exist in
atomic $\Lambda$ systems has previously been considered
\cite{Rybin2004,Konopnicki1981}

In conclusion, we have demonstrated the feasibility of an
alternative retrieval scheme for dynamic EIT under certain
parameter regimes and provided physical explanations for its
behavior. Our numerical simulation also demonstrated the ability
of this new scheme to create soliton-like features that are
sensitive to the relative coupling strength of the two
transitions. Due to its sensitivity to parameter change, we
believe this solitonic behavior could prove useful within the
context of magnetometry or high precision measurement.

We thank Elena Ostrovskaya for helpful discussions regarding the
soliton features.

\end{document}